\documentclass[aps,twocolumn,showpacs,superscriptaddress,floatfix]{revtex4}

\usepackage{amssymb}
\usepackage{epsfig}
\usepackage{amsmath}
\usepackage{graphicx}
\usepackage{rotating}
\usepackage{color}

\begin{document}

\title{On detecting Higgs coupling in transitions of light atoms}

\author{Rajmund Krivec}

\affiliation{Department of Theoretical Physics, J. Stefan Institute,
Jamova 39, 1000 Ljubljana, Slovenia}

\email{rajmund.krivec@ijs.si}



\begin{abstract}
In light of the known Higgs mass and the current constraints on the
quark-lepton Higgs coupling, we derive conditions for extracting upper
limits on the lepton-nucleon Higgs coupling from light atoms and ions,
assuming the availability of locally precise two- and three-body methods
might be beneficial. A recent work has proposed to extract these limits in
heavy atoms where the Higgs term is enhanced by $\approx 10^3 AZ$, due to
both the large coupling modifier and large $A$, $Z$, and assuming
sufficiently precise relativistic electron wave functions. We first revisit
the old idea of using the Lamb shift in light muonic ions where the coupling
is enhanced by about $201^3 AZ^3$ primarily due to the concentration of the
muon wave function at the origin, the muon coupling modifier already being
close to 1. For the muonic helium an experimental precision below 0.1 ppm is
required to reach the constraints on Higgs couplings. However, theoretical
uncertainty is large due to nuclear potential dependence of the finite size
terms enhanced by the small muon orbit, and their elimination by using
several states is precluded due to the Lamb shift being the only precisely
measurable state. In normal (electronic) light systems transitions between
low-lying states lie near the optical region allowing precise experiments,
and extraction may be possible by eliminating the finite-size, polarization
and Zemach moment terms from a set of transitions, e.g. $1S-2S$ and improved
$2^3S-2^3P$ and $2^1S-2^3S$ in ${\rm He}^+$, while isotope shifts could be
used if additional transitions are measured as precisely.
\end{abstract}

\maketitle

\section{Introduction}

There has been a proposal to extract limits on Higgs nucleon-lepton coupling
constant from valence electron transitions in heavy atoms \cite{P1,P1a,P2}.
In the present paper we examine the conditions for extracting these limits
from transitions in light muonic or normal atoms or ions. The advantages and
disadvantages of the two proposals are as follows.

The heavy-atom proposal \cite{P1,P1a,P2} is based on the coupling
enhancement due to the large atomic number $A$; the stability of systems
with large $A$ allowing the experiment to reach precision of the order of
$1$ Hz in atomic clock transitions; and the relatively small other
corrections like the weak force, despite the fact that it may mask the Higgs
contribution. Due to the large number of precisely measurable transitions
available, uncertainties in theoretical corrections depending on total
charge $Z$ can be conveniently avoided \cite{P1,P1a} by the use of isotope
shifts, i.e. the deviations from linearity in the King's plots \cite{K1}, as
the change in $A$ affects the transitions via nuclear recoil, electron
correlations and nuclear charge radius independently of the transition
measured. Disadvantages are the approximate nature of the factorization of
the screened, relativistic electron wave function entering the transition
matrix element, and the reliance on the existence of new physics via the
relatively large current value (of the order of $10^3$) of the coupling
modifier $\kappa_e$ for the electron-Higgs coupling constant $y_e$ relative
to its Standard Model (SM) value, $y_e = \kappa_e y_e^{\rm SM}$.

The light-system proposal seems attractive as even for the three-body system
$e\mu^4{\rm He}$ locally precise nonrelativistic wave functions can be
calculated by e.g.\ the Correlation-function Hyperspherical Harmonic Method
(CFHHM) \cite{KMmuHe}, while for two-body ions like $\mu^4{\rm He}^+$ both
relativistic and non-perturbative methods are available. (In heavier muonic
systems, electron screening of muonic orbits remains weak but scaling of
nuclear structure effects strongly amplifies theoretical uncertainties.)

We start by applying the known Higgs mass to the old estimates \cite{B1},
which used a much smaller mass, of the Higgs contribution to the muonic
$^4{\rm He}$ Lamb shift. This is measured by the CREMA collaboration
\cite{CREMA} to a $10^{-5}$ level using laser spectroscopy. Different muon
and electron reduced masses resulting in small muonic orbits provide a large
coupling enhancement $(m_{\mu A}/m_{eA})^3$. A disadvantage is the $2S$
state lifetime of the order of 1 $\mu{\rm s}$ \cite{Lamb2012,CREMA2016}, a
result of the finite muon lifetime of 2.2 $\mu$s, the $1S-2S$ two photon
decay time of 8 $\mu{\rm s}$ and the collisional quenching rate in gas. It
prevents detecting effects below $1$ MHz ($10^{-6}$ meV) and complicates
preparation of states. Another disadvantage is that while the sum of QED
corrections ($1813.02$ meV for $\mu^4{\rm He}^+$ \cite{B1}) is well known
and can be improved, the finite nuclear size and polarization corrections
also scale with the lepton mass, amplifying the uncertainties in the nuclear
charge radii.

Next we look at normal (electronic) light systems where the precisely
measured transitions (near the optical region) are those between low-lying
$nS$ states, losing the enhancement by the muon but benefitting from smaller
nuclear structure corrections and measurements at the $10^{-12}-10^{-15}$
level, related to the extraction of electronic charge radii \cite{CREMA} and
the Rydberg constant.

A common disadvantage of light systems is that the perturbation theory in
$(\alpha Z)$, where $\alpha$ is the fine structure constant, has been well
studied only up to $(\alpha Z)^6$, as appropriate for charge radii
extraction \cite{FriarFS,B2}. For muonic systems the nonrelativistic and
relativistic expansions give identical results \cite{FriarFS} while for
electronic systems relativistic treatment is required
\cite{Yerokhin2011,Czarnecki2005}.

In the next section we calculate the current orders of magnitude of the
Higgs coupling parameters for light systems. This is followed by an overview
of the current light ion physics and the bottlenecks for reducing its
uncertainty. The last section gives requirements on theoretical terms
appearing in the expression for the coupling constant based on the isotope
shifts when only two transitions are available.

\section{Bounds on the Higgs term}

Higgs exchange between a nucleus and a bound electron or muon results in a
potential of the Yukawa type,

\begin{equation}
V_{H}(r) = - g_{H\mu A} \frac{e^{-m_{H}r}}{r},
\end{equation}
where $g_{H\mu A}$ is proportional to the muon and nuclear coupling
constants,
\begin{equation}
g_{H\mu A} = \frac{y_\mu y_A}{4\pi}.
\end{equation}
The SM fermion-Higgs coupling constants are proportional to the fermion
($F$) mass according to the assumed hierarchy leading to fermion masses, as
well as to the coupling modifiers based on experimental upper bounds on
couplings which allow for the existence of new physics:
\begin{equation}
y_F = \kappa_F \frac{m_F}{v}, \ v = 246\, {\rm GeV}.
\end{equation}

For the electron,
\begin{equation}
y_e = \kappa_e \times 2.1 \times 10^{-6}.
\end{equation}
Ref.\ \cite{P1} uses $y_e$ at the upper bound set by the LHC data on
${H}\rightarrow e^+e^-$ \cite{Ye1,Ye2,Ye3}, where $\kappa_e < 611$
\cite{Ye3},
\begin{equation}
y_e < 611 \times 2.1 \times 10^{-6} \approx 1.3 \times 10^{-3}.
\end{equation}
This $\kappa_e$ value corresponds to the lowest new physics scale, $5.8$ TeV
\cite{Ye3}, but the bounds are likely to improve, lowering the $y_e$ value
and reducing the feasibility of the proposal \cite{P1,P1a}.

For the muon, $\kappa_\mu = 0.2_{-0.2}^{+1.2}$ in Table 15 of Ref.\
\cite{kmu1} so we do not get the advantage of a weak experimental upper
bound, resulting practically in the SM coupling value,
\begin{equation}
y_\mu \lesssim 1.4 \times 207 \times y_e^{\rm SM} \approx 0.6 \times 10^{-3},
\end{equation}
using the upper bound of $\kappa_\mu$.

The nuclear coupling is approximately proportional to the atomic number $A$,
\begin{equation}
y_A = (A - Z) y_n + Z y_p \approx A\, y_N,
\end{equation}
where $y_n \approx y_p \approx y_N$ are the neutron and proton coupling
constants which are linear combinations of the quark and gluon coupling
constants. In more detail \cite{Yq1,Yq2,Yq3,Yq4,P1} and neglecting the $c_g$
term \cite{Yq5},
\begin{eqnarray}
y_n\approx 7.7 y_u + 9.4 y_d + 0.75 y_s, \nonumber \\
y_p\approx  11 y_u + 6.5 y_d + 0.75 y_s.
\end{eqnarray}
The weakest bounds on individual quark couplings are $y_q \lesssim 0.3$
\cite{Yudsc1,Yudsc2,Yudsc3}, where $y_q$ is one of $y_u$, $y_d$, $y_s$, or
$y_c$, resulting in $y_N \lesssim 3$ due to suppression of light quarks. LHC
and electroweak data give a medium bound $y_q \lesssim 1.6 \times 10^{-2}$
\cite{Yudsc2,Yudsc4,Yudsc5}, resulting in $y_N \lesssim 0.2$. Indirect
bounds may be even lower, $y_q \lesssim 5 \times 10^{-3}$ \cite{P1,Yudsc6}
resulting in $y_N \lesssim 10^{-3}$. These results translate to the
following range of current upper bounds on the nuclear coupling $y_A$ in muonic
hydrogen,
\begin{equation}
y_1 \approx \{10^{-3},\, 0.2,\, 3\},
\end{equation}
and in muonic $^4{\rm He}$:
\begin{equation}
y_4 \approx 4 \times \{10^{-3},\, 0.2,\, 3\}.
\end{equation}

The corresponding bounds on $g_{H\mu A}$ are linear in $A$:
\begin{equation}
g_{H\mu 1} \lesssim \{ 0.5\times10^{-7},\, 1\times10^{-5},\, 0.1\times10^{-3}\}
\label{eqgh1}
\end{equation}
for $\mu {\rm H}$ and
\begin{equation}
g_{H\mu 4} \lesssim \{ 2\times10^{-7},\, 4\times10^{-5},\, 0.6\times10^{-3}\}
\end{equation}
for $\mu^4{\rm He}^+$/$e\mu^4{\rm He}$.

Due to the large Higgs mass, the Higgs term is given by the leading order of
perturbation in $(1/m_{H})^2$, unlike Ref.\ \cite{B1} where $m_{H}$ ranged
from $0.15$ MeV to $750$ MeV. It is negligible in states with orbital
angular momentum $l > 0$. For the Lamb shift, at the principal quantum
number $n = 2$, the Higgs term in the transition energy $\Delta E_{LS}$
comes from the $2S$ state matrix element,
\begin{equation}
\delta_H(\Delta E_{LS}) \approx g_{H\mu A} \bigl\vert
R_{20}(0)\bigr\vert^2 \frac{1}{m_{H}^2}, \label{eqM}
\end{equation}
where $R_{20}$ is the $n=2$, $l=0$ radial wave function $R_{nl}$ of the
lepton $x$,
\begin{equation}
\bigl\vert R_{n0}(0)\bigr\vert^2
    = 4 \Bigl( \frac{\alpha Z m_{xA}}{n} \Bigr)^3.
\label{eqR}
\end{equation}
In general, for transition $i$,
\begin{equation}
\delta_H(\Delta E_i) \approx
g_{HxA} \ \frac{4}{m_{H}^2} \Bigl(\alpha Z m_{xA}\Bigr)^3 \Delta_i ,
\label{eqRx}
\end{equation}
where $\Delta_i = \delta_{l_{i1,0}}/n_{i1}^3 - \delta_{l_{i2,0}}/n_{i2}^3$.

The $201^3$-fold enhancement of the muonic Higgs term relative to electronic
transitions at the same $n$ due to the smaller muon reduced Bohr radius
together with the $\approx AZ^3$ scaling results in total enhancement
$\approx 201^3 \times AZ^3$.

For comparison, the coupling in the proposal \cite{P1,P1a,P2} is enhanced
relative to single nucleon-electron coupling through $y_A$ by way of large
$A$, and through the electron wave function squared at origin, $\vert
\psi_e(0)\vert^2$, by way of large $Z$ (screening in heavy atoms precludes
enhancement by $Z^3$ by an unperturbed electron wave function, resulting in
enhancement by $Z$ instead \cite{P1}). At the current coupling modifier
their total enhancement is $\approx 611 \times AZ$.

\begin{table}[h]
\begin{center}
\caption{Requirements for extracting the Higgs contribution $\delta_H(\Delta
E_{LS})$ from the Lamb shift of $\mu^4{\rm He}^+$ and $\mu{\rm H}$ for the
known Higgs mass (except row 5, see below). $\delta_H\nu$ are the
corresponding frequencies. Bounds on the nucleon-Higgs coupling $y_A$ are
from Ref.\ \cite{P1}. $\eta_H = 10^6 \vert\delta_H(\Delta E_{LS})/\Delta
E_{LS}\vert$ is the required precision in ppm. Rows 4 and 5 correspond to
the precision defined as the discrepancy between theory and experiment in
Ref.\ \cite{B1}: row 4 shows the extrapolated $g_{H\mu 4}$ needed for
observation of $125$ GeV Higgs at that $\eta_H$, and row 5 is for an old 200
ppm experiment using the Higgs mass $0.75$ GeV as dicussed in Ref. \cite{B1}
alos for lower masses. The $\mu{\rm H}$ result (row 6) is for upper bound on
Higgs coupling.}
\begin{tabular}{cr|c|cc|c}
\hline
system
& $\Delta E_{LS}$ & $g_{H\mu A}$       &  $\delta_H(\Delta E_{LS})$  & $\delta_H\nu$ & $\eta_H$ \\
&  (meV)   &                    &        (meV)                & (Hz)          & (ppm) \\
\hline
\hline
$\mu^4{\rm He}^+$
&$1664$ & $2  \times10^{-7}$   & $2  \times 10^{-8}$ & $5 \times 10^3$ & $10^{-5}$ \\
&       & $4  \times10^{-5}$   & $4  \times 10^{-6}$ & $1 \times 10^6$ & $3\times10^{-3}$ \\
&       & $0.6\times10^{-3}$   & $0.6\times 10^{-4}$ & $2 \times 10^7$ & $< 0.1$ \\
\hline

&       & $36$                 & $4                 $ & $1.0 \times 10^{12}$ & $3\times10^3$\\

&       & $1.3 \times 10^{-3}$ & $4                 $ & $1.0 \times 10^{12}$ & $3\times10^3$\\
\hline
$\mu{\rm H}$
&$202$  & $0.1\times10^{-3}$   & $1  \times 10^{-6}$ & $3\times 10^5$ & $< 0.01$ \\
\hline
\end{tabular}
\label{higgs1}
\end{center}
\end{table}

Table \ref{higgs1} gives experimental accuracy requirements for extracting
the Higgs term from Lamb shift for $\mu^4{\rm He}^+$ or $e\mu^4{\rm He}$ and
$\mu{\rm H}$. Errors due to uncertainties of the fundamental constants are
of the order of $10^{-5}$ meV \cite{Martynenko07}. The Higgs term lies above
the muon decay limit and within the upper range of $g_{H\mu 4}$ only in
muonic helium or heavier systems. The $Z^3$ transition energy scaling also
favors heavier systems as the required relative accuracy is smaller.


For fixed coupling modifiers, the Higgs coupling scales to the normal
(electronic) light systems as $g_{He A} \approx 2.2\, g_{H\mu A}$ for the
same $A$, $n$, and the Higgs term scales as $g_{He A} \, (m_{eA}/m_{\mu
A})^3 \approx 0.27\times 10^{-6}$.

\begin{table}[h]
\begin{center}
\caption{As in Table \ref{higgs1}, but for the normal (electronic) systems
and for the corresponding upper bounds (proportional to $A$) on the Higgs
coupling; relative accuracy is omitted.}
\begin{tabular}{ccc|c|cc}
\hline
system         & transition
 & $\Delta E$       & $g_{He A}$ &  $\delta_H(\Delta E)$      & $\delta_H\nu$ \\
               &
 & (meV)            &            &  (meV)                     & (Hz)          \\
\hline
\hline
$^4{\rm He}^+$ & Lamb shift
 & $0.05$           & $1.3\times10^{-3}$   & $2 \times 10^{-11}$ & $4$   \\
               & $2^3S-2^3P$
 & $1.15\times10^3$ &                      & $2 \times 10^{-11}$ & $4$   \\
               & $1S-2S$
 & $40\times10^3$   &                      & $1 \times 10^{-10}$ & $30$  \\
H              & $1S-2S$
 & $10\times10^3$   & $3  \times10^{-4}$   & $3 \times 10^{-12}$ & $0.8$ \\
\hline
\end{tabular}
\label{electron1}
\end{center}
\end{table}


In electronic systems the Lamb shift is not in the optical range and not
very precisely measured. In the helium ion it is $14\,{\rm GHz} \pm\,
348\,{\rm kHz}$ or $0.05$ meV \cite{Herrmann2009} (Table \ref{electron1}),
the Higgs term is $4$ Hz ($1.7 \times 10^{-11}$ meV) at saturated coupling,
and the uncertainty would have to be decreased by $10^5$, more than in
muonic systems. Better precision is achieved in the near-optical
transitions, the centroid $2^3S-2^3P$ transition having $2.4$ kHz
uncertainty ($10^{-8}$ meV) \cite{PastorHeIS2012}, and in the $1S-2S$
transition \cite{Herrmann2009}. For comparison we give the hydrogen $1S-2S$
transition \cite{CREMA}.

The current uncertainities in the above transitions and the required
increase in precision is discussed in the next section.

\section{Current uncertainties}

Experiments in light systems focus on the extraction the of charge radii and
the Rydberg constant. We identify suitable transitions for Higgs term
extraction and the bottlenecks for reducing their uncertainties.

The $2S_{1/2}-2P_{3/2}$ and $2S_{1/2}-2P_{1/2}$ transitions used for
calculating \cite{Carlson2011,Carlson2015} the $\mu^4{\rm He}^+$ Lamb shift
were measured long ago at about $1528$ meV and $1381$ meV, respectively
\cite{Carboni77,Carboni78}. At the time of the proposal \cite{B1} based on
the light Higgs, the discrepancy between theory and experiment as given by
the uncertainty of the finite-size ($-288.9 \pm 4.1$ meV) and nuclear
polarization terms ($3.1\pm 0.6$ meV) was an order of magnitude larger than
the experimental uncertainties, $\pm 0.3$ meV ($\pm 0.5$ meV) or 200 ppm
(330 ppm), respectively. (The electron scattering $^4{\rm He}$ radius used
was $1.674\pm0.012$ fm \cite{Sick76}.)

Improved calculations \cite{Martynenko07} reduced the uncertainty of the
non-nuclear contributions to the Lamb shift to $10^{-3}$ meV (240 MHz, 0.6
ppm) for $\mu ^4{\rm He}^+$, but that of the finite-size correction
($-295.848\pm 2.8$ meV) was still large (relative error $1\times 10^{-2}$),
double the error of the charge radius \cite{Martynenko07}, and that of the
nuclear polarization term of the two-photon exchange correction remained
$0.6$ meV. The uncertainty of the $^4{\rm He}$ charge radius was $1.676(8)$
fm ($5\times 10^{-3}$ relative error) \cite{Friar03,Sick82}. The charge
radius puzzle in the proton \cite{PRP1,PMP1,Bernauer1} spurred new
measurements; it has recently been confirmed in $\mu{\rm D}$ \cite{DRP1}.
The electron scattering $^4{\rm He}$ charge radius is known to
$2\times10^{-3}$ accuracy ($1.681 \pm 0.004$ fm)
\cite{AntogniniPRP1,SickRe4he}.


As the current Lamb shift experiments aim to resolve the charge radius
problem \cite{CREMA,Pohl2016a}, theoretical work is dedicated solely to
improving the polarization terms but not the finit-size terms.
(The complementary measurements of the electronic $1S-2S$ transition in
$^4{\rm He}^+$ serve to test the QED part \cite{Herrmann2009}.) Laser
spectroscopy of $\mu^3{\rm He}^+$ and $\mu^4{\rm He}^+$ at ``moderate''
precision of 50 ppm \cite{Lamb2012,AntogniniPRP1} yielded the charge radius
to $1\times 10^{-3}$ accuracy \cite{Lamb2012}. The 2013-2014 $\mu^3{\rm
He}^+$ and $\mu^4{\rm He}^+$ 40 ppm measurements
\cite{CREMA2016,Antognini15} yielded the $^4{\rm He}$ and $^3{\rm He}$
charge radii to about $3\times10^{-4}$ \cite{CREMA2016}.

The transitions actually measured for the Lamb shift \cite{Carlson2015} are
$2S_{1/2}-2P_{3/2}$ and $2S_{1/2}-2P_{1/2}$ in $\mu ^4{\rm He}^+$; the six
transitions between the HFS-split $2S$ and $2P$ states planned in Ref.\
\cite{Lamb2012,AntogniniPRP1} in $\mu ^3{\rm He}^+$; and the
$2S_{1/2}^{F=1}-2P_{3/2}^{F=2}$ and $2S_{1/2}^{F=0}-2P_{3/2}^{F=1}$
\cite{CREMA2016,Carlson2011,Carlson2015} in $\mu {\rm H}$. These are
combined with the theoretical $2P_{3/2}-2P_{1/2}$ fine splittings and with
the hyperfine splitting (HFS) of the ground state for nonzero-spin nuclei,
or the electron-muon HFS for high precision \cite{KMmuHe}, The latter also
allows the determination of the Zemach radius from the nuclear polarization
or vice versa. The HFS precision, $4465.004(29)$ MHz ($1.8\times10^{-3}$
meV), was 6.5 ppm \cite{Gardner82}, while the CREMA collaboration is aiming
at 1 ppm \cite{CREMA2016}. (These experiments are also used to check some
terms in the Lamb shift: a contribution from the nuclear polarization term
cancels the third Zemach moment $\langle r^3\rangle_{(2)}$ in $\mu^4{\rm
He}^+$ \cite{Barnea13a} like it was observed earlier for $\mu{\rm D}$
\cite{FriarZem,PachuckiZem}.) The current results are reviewed in Ref.\
\cite{B2} up to $\mu^4{\rm He}^+$ and in Ref.\ \cite{PohlSOP} for $\mu {\rm
H}$. The vacuum polarization (VP) term contains the following uncertainties
\cite{B2,Karshenboim2010,Karshenboim2012,Jentschura2011b,Elekina2011,Krutov2015}.
The relativistic perturbative Uehling term in $\mu {\rm H}$ is modified by
finite-size effects by about $0.0079$ meV and $0.0082$ meV for the two
proton radii involved in the proton radius problem, $0.842$ fm and $0.875$
fm, respectively. The $\mu^4{\rm He}^+$ finite-size effect in VP is
$-0.3297\,\langle r_\alpha^2\rangle$ meV fm$^{-2}$ implying a similar
uncertainty of $0.0016$ meV for the current $5 \times 10^{-3}$ $^4{\rm He}$
radius uncertainty. Neglecting finite nuclear size in muon-electron VP
causes moderately increasing shifts of up to $0.0001$ meV for $\mu^4{\rm
He}^+$. Uncertainties of the ``light-by-light'' corrections reach $0.0006$
meV for $\mu^4{\rm He}^+$, while the sixth-order VP uncertainties reach
$0.003$ meV. The largest uncertainty within the VP terms is the hadronic VP,
reaching an estimated 5\% uncertainty in the $0.225$ meV value for
$\mu^4{\rm He}^+$, or estimated $0.012$ meV \cite{B2}. The relativistic
recoil amounts to about $0.001$ meV
\cite{Karshenboim2010,Karshenboim2012,Jentschura2011b,Elekina2011,Krutov2015}.

To return to the nuclear structure terms, the finite-size term
proportional to the charge radius squared $\langle r_p^2 \rangle$ in $\mu
{\rm H}$ has uncertainty $0.064$ meV for the spectroscopic radius $0.875$ fm
and $0.010$ meV for the Lamb shift radius $0.842$ fm; however, the
corresponding values $-3.978$ meV and $-3.6855$ meV differ by 8\%, or $0.3$
meV. The situation in $\mu^4{\rm He}^+$ is worse due to the scaled
contribution, amounting to $1.4 - 2.8$ meV uncertainty depending on which
radius is taken \cite{B2,Martynenko07}.

The Lamb shift is usually parametrized in terms of charge distribution
moments as ${\cal A} + {\cal B}\langle r^2\rangle + {\cal C}(\langle
r^2\rangle)^{3/2}$ which is suitable at current precision; it also has the
consequence that the measured Lamb shift itself is rarely quoted
\cite{DiepoldPhD}. For $\mu^4{\rm He}^+$, ${\cal B} = -106.344$ meV
fm$^{-2}$ \cite{B2}, and consists of six contributions, the largest being
the leading term
\begin{equation}
b_a
= -\frac{2\alpha Z}{3} \Bigl(\frac{\alpha Z m_{\mu A}}{n}\Bigr)^3
\label{eqba}
\end{equation}
in Eq.\ (5) of Ref.\ \cite{FriarFS}, amounting to $-105.319\,\langle
r_\alpha^2\rangle$ meV fm$^{-2}$. (The total includes the finite-size VP
correction $-0.3297\,\langle r_\alpha^2\rangle$ meV fm$^{-2}$ quoted above.)
In $^4{\rm He}^+$, the nonleading contributions $b_b + {\ldots} + b_e$
amount to about a percent. Some depend on the assumed analytic charge
distribution via terms e.g.\ $b_a (\alpha Z)^2 \langle \ln(\alpha Z m_{\mu
A} r)\rangle$ (Ref.\ \cite{B2}, Appendix B, and Ref.\ \cite{FriarFS}). $\cal
C$ involves a model-dependent transformation between the third Zemach moment
$\langle r^3\rangle_{(2)}$ and $(\langle r^2\rangle)^{3/2}$ via a factor
$f_{Zem}$ which for $\mu ^4{\rm He}$ is about $3.5$, but depends on the
charge distribution already on the second digit \cite{B2}. In $\mu ^4{\rm
He}$ the uncertainty of the ${\cal C}$ term is one-third the uncertainty of
the polarization term.
The relativistic corrections start at $(\alpha Z)^6$ as
verified in Ref.\ \cite{FriarFS} using perturbation theory based on both the
Schr\"{o}dinger and the Dirac wave functions. Radius-independent corrections
for $\mu {\rm H}$ are summarized in Table 1 of Ref.\ \cite{PohlSOP}.

Reduction of the charge radius uncertainty to the level of the parameter
dependence of the effective nuclear potentials has been achieved in recent
{\sl ab-initio} calculations of the inelastic term in the two-photon
exchange correction in light muonic atoms using state of the art nuclear
potentials (AV18 and $\chi$EFT) and the hyperspherical harmonic EIHH method
\cite{Barnea13a,Barnea15b,Barnea16b}. The nuclear problem was solved
separately and the polarization terms evaluated in second-order perturbation
theory in terms of the residual Coulomb potential for point nucleons. A
$5\times 10^{-2}$ accuracy is required for determining the $^3{\rm He}$ and
$^4{\rm He}$ charge radii squared to $3\times 10^{-4}$
\cite{Barnea13a,Barnea15b,CREMA2016,AntogniniPRP1}, ensuring the same
absolute errors in both terms. The new value of the nuclear polarization
term $-2.47(14)$ meV \cite{Barnea13a} has $6\times 10^{-2}$ accuracy
(absolute error is misquoted as $0.015$ meV in Ref.\ \cite{B2}) compared
with the old \cite{B1} value $3.1\pm 0.6$ meV ($2\times 10^{-1}$ accuracy).
The AV18 and $\chi$EFT potentials are tuned to the $^3{\rm He}$ binding
energy but they give different charge radii. Uncertainty of the polarization
term may be further reduced using the $^4{\rm He}$ charge radius to
constrain the nuclear potential models \cite{Barnea13a}, but beyond that any
improvement seems unlikely.

\begin{table}[h]
\begin{center}
\caption{Coefficients of the $\mu {\rm H}$ Lamb shift parametrization ${\cal
A} +
{\cal B}\langle r_p^2\rangle + {\cal C}(\langle r_p^2\rangle)^{3/2}$ for perturbative
\cite{B2} and nonperturbative calculations. Higher terms from Ref.\
\cite{muHNonPert2013} are not quoted.}
\begin{tabular}{c|lll}
\hline
Ref.               & ${\cal A}$ (meV) & ${\cal B}$ (meV fm$^{-2}$) & ${\cal C}$ (meV fm$^{-3/2}$) \\
\hline
\hline
\cite{B2}             & 206.0611(60) & -5.22718             & 0.0365(18) \\
\hline
\cite{muHNonPert2011} & 206.0604     & -5.2794              & 0.0546 \\
\cite{muHNonPert2013} & 206.0465137  & -5.226988356         & 0.03530609322 \\
\hline
\end{tabular}
\label{higgs3}
\end{center}
\end{table}

Current nonperturbative calculations also suffer from the nuclear model
dependence via the assumed charge distributions. They have been performed
for muonic hydrogen (and could be extended to helium). The work
\cite{muHNonPert2011} solves the Dirac equation to $500$ neV accuracy but
describes recoil only via the reduced mass of the muon leaving further
corrections to perturbation theory. The charge distributions were given in
terms of moments, i.e.\ $\langle r_p^2 \rangle$, to express results in the
conventional form. Also, a number of corrections were not calculated
\cite{B2} (two- and three-loop VP, muon self-energy, muon and hadron VP, and
nuclear polarization). The terms differ from the perturbation theory to the
order of $0.03$ meV. The work \cite{muHNonPert2013} using the proton dipole
form factor, Gaussian, uniform, Fermi and experimentally fitted charge
distributions seems better converged, listing the calculated terms to better
than $0.001$ meV accuracy, but the charge distribution dependence was of the
order of $0.004$ meV in the Coulomb and VP terms. Methods are compared in
Table \ref{higgs3}, with the differences up $0.05$ meV.

In summary, the uncertainty of the nuclear structure terms in light muonic
systems is $3-4$ orders of magnitude larger than the requirements in Table
\ref{higgs1} and cannot be reduced further. The Zemach moment term has about
$0.2$ meV uncertainty in muonic helium \cite{B2}.

In the normal (electronic) light systems the precision is higher and
corrections must be based on the Dirac wave functions. The extensive
literature \cite{Beauvoir2000, Herrmann2009, Yerokhin2011, PastorHeIS2012,
PachuckiHeIS2015, CREMA} is reviewed e.g. in CODATA \cite{CODATA2012}.

The highest precision is achieved in H which is less favorable for Higgs
term extraction than He (Table \ref{electron1}). The $1S-2S$ transition in
hydrogen used for deducing the Rydberg constant is currently measurable with
$4\times 10^{-15}$ ($1$ Hz) uncertainty \cite{CREMA}. This almost meets the
requirement of Table \ref{electron1}, but it is overshadowed by the
uncertainty of the nuclear structure corrections, as the relative size of
the nuclear contributions to transition energy is $4\times10^{-10}$, or
about 1 MHz, therefore they should be known to better than 6 places for
direct extraction of the Higgs term, clearly not achievable as the charge
radius uncertainty (a decade ago) was 44 kHz and the $B_{60}$ and $B_{7i}$
terms of the two-loop QED corrections \cite{Yerokhin2006} are $-8$ kHz
\cite{Herrmann2009}.

In $^4{\rm He}^+$, The $1S-2S$ transition at $9.9\times 10^{16}$ Hz is
already predictable with $0.35$ MHz uncertainty which is 4 orders of
magnitude worse that the requirement of Table \ref{electron1}
\cite{Herrmann2009}, the largest uncertainties stemming from the charge
radius and the $B_{60}$ and $B_{7i}$ terms. The accuracy of the $1S-2S$
transition could be improved to $10^{-16}$ ($1$ Hz) \cite{Herrmann2009},
which would be 30 times better than the requirement of Table
\ref{electron1}. Other suitable transitions, for example the $2^3S-2^3P$ at
$1.1$ eV, are currently at the $10^{-10}$ uncertainty or $2$ kHz level
\cite{PastorHeIS2012}, which is 3 orders of magnitude short of requirements
of Table \ref{electron1}. The nuclear structure contributions are amplified
with respect to hydrogen. They are given in detail in Ref.\
\cite{PachuckiHeIS2015}. The nonlogarithmic relativistic correction $f_{\rm
fs}$ depends on the assumed nuclear charge distribution model in the leading
digit, and its relative contribution $(Z\alpha)^2 f_{\rm fs}$ to the finite
nuclear size term is $4\times10^{-5}$.

In transitions between $2S$ and $2P$ states of the normal (electronic)
helium the finite-size terms, which scale as $Z^4 m_{xA}^3$, can be
estimated to be at the $0.4\times10^{-4}$ meV ($9$ MHz) level so direct
extraction of the Higgs terms is ruled out here as well. The Zemach moment
term is $4\times10^{-5}$ times the finite-size term \cite{PachuckiHeIS2015},
or $1.6 \times 10^{-9}$ meV, so a typical one percent nuclear uncertainty
would appear at the $10^{-11}$ meV level. This is close to the requirement
of Table \ref{electron1} so this term cannot be a priori excluded. (The
uncertainty in the Rydberg constant, known to $2\times10^{-11}$
\cite{CREMA}, cancels out in the Higgs extraction.)

\section{Extraction of the Higgs term}

In normal heavy atoms of proposal \cite{P1,P1a,P2} isotope shifts are the
most promising method of extracting the Higgs term, by looking for the
departure from the linearity of the King's plots \cite{P1a,K1} for a pair of
transitions. For large $A$, $A'$, the coefficient of $\langle r^2\rangle_{A}
- \langle r^2\rangle_{A'}$ in the $A - A'$ isotope shift is essentially
independent of $A$, $A'$. For example, the relative isotope shift of
electron reduced masses for $A=100$, $A'=101$ is $5\times 10^{-8}$ and that
of the leading term of the finite-size coeffficient is three times that. A
similar argument regarding the King's plot linearity is made in Ref.\
\cite{P1}. Using isotope shifts of two measured transitions we can eliminate
the $\langle r^2\rangle_{A} - \langle r^2\rangle_{A'}$ terms. The validity
across several isotope pairs $A$, $A'$ of the resulting linear relation
between the two isotope shifts can then be studied. This requires at least
two transitions measured for three different $A$, but there are many
suitable transitions in heavy atoms.

In light systems we have a similar parametrization of transition energies
but with the additional term (Zemach moment term) ${\cal C}'\langle
r^3\rangle_{(2)}$ that needs to be eliminated together with ${\cal B}\langle
r^2\rangle$ (the Zemach moment has to be used to avoid the model-dependent
transformation factor $f_{Zem}$ \cite{B2, PachuckiHeIS2015}). To reduce the
number of required transitions, one could presumably do this without isotope
shifts using three transitions ($i = 1, 2, 3$) of a fixed isotope, provided
${\cal A}_i$, ${\cal B}_i$, ${\cal C}'_i$ are known:
\begin{eqnarray}
\Delta E_i =
{\cal A}_i + {\cal B}_i \langle r^2\rangle +
{\cal C}'_i \langle r^3\rangle_{(2)}  + c A{\cal H}_i
\label{eqdei}
\end{eqnarray}
where
\begin{eqnarray}
c = g_{H\mu 1} = \frac{y_\mu y_N}{4\pi} < 3\times 10^{-4}, \\
{\cal H}_i = \frac{4}{m_H^2} \Bigl(\alpha Z m_{xA}\Bigr)^3 \Delta_i.
\end{eqnarray}
(We leave out the weak interaction term \cite{P1,P1a}.) Assuming for
simplicity that $\Delta_3 = 0$ (see below),
\begin{equation}
c=\frac{1}{A}
\frac
{( e_1 {\cal B}_{23} - e_2 {\cal B}_{13} ) -
 ( {\cal C}'_1 {\cal B}_2 - {\cal C}'_2 {\cal B}_1 ) {e_3}/{{\cal C}'_3}}
{{\cal H}_1 {\cal B}_{23} - {\cal H}_2 {\cal B}_{13}}
\end{equation}
where
\begin{equation}
e_i=E_i-{\cal A}_i, \ \ 
{\cal B}_{ij} = {\cal B}_i - \frac{{\cal C}'_i}{{\cal C}'_j} {\cal B}_j.
\end{equation}
In this case measurements on a different isotope $A'$ if available would be
used independently to improve $c$.

Isotope shifts in light systems introduce yet more terms requiring more
transitions to eliminate them. The isotope shift of the leading contribution
$b_a$ (Eq.\ (\ref{eqba})) to ${\cal B}$ is $3$ percent for $A=3$, $A'=4$ due
to muon reduced mass shift. $b_b$, $b_c, {\ldots} $ \cite{B2} also depend on
$A$ via $m_{\mu A}$ starting at order $(\alpha Z)^6$. In electronic light
systems the relative isotope shift of ${\cal B}$ is $10^{-4}$ (negligible
\cite{B2} for the charge radius determination but not for the Higgs
extraction). If we denote the isotope shift of $a$ by $[a]_{AA'} = a_A -
a_{A'}$, we have for the transition energy $\Delta E_i$ (leaving out the
$AA'$ suffix for brevity):
\begin{eqnarray}
[\Delta E_i] =
[{\cal A}_i] +
&&{\cal B}_i  [\langle r^2\rangle]       + [{\cal B}_i] \langle r^2
\rangle_{A'} +~~~~\\ \nonumber
&&{\cal C}'_i [\langle r^2\rangle^{3/2}] + [{\cal C}'_i] {\langle r^3\rangle_{(2)} }_{A'} +
c[A{\cal H}_i].~~~
\label{eqdeiis}
\end{eqnarray}
The term corresponding to $[{\cal B}_i]_{AA'} \langle r^2 \rangle_{A'}$ is
claimed sufficiently small for heavy atoms \cite{P1}. The Zemach moment term
($[{\cal C}'_i]_{AA'} {\langle r^3\rangle_{(2)}}_{A'}$) may also turn out to
be small enough (e.g. in $^4{\rm He}^+$, above), permitting the use of four
instead of five transitions. Obviously we cannot look for King's plot-type
linearity here as we are likely to have only 2 isotopes. Also, and the
coefficients ${\cal B}_i$ and ${\cal C}'_i$ may depend on $A$ appreciably in
higher orders of $(\alpha Z)$.

In $\mu{\rm He}$, even if the precision of the Lamb shift proper could be
improved (Table \ref{higgs1}), there is no suitable second transition; the
$1S-2S$ (or the appropriate centroid energy \cite{PachuckiHeIS2015,
Yerokhin2011}) at 8 keV lies in the X-ray region where the experimental
precision is smaller but the required relative precision is $10^3$ times
larger than for the Lamb shift. (We cannot take the two transitions to be a
pair of the separate transitions measured for the Lamb shift as their Higgs
terms cancel and, for isotope shifts, quantum numbers have no counterparts
beteen $A=4$, $A'=3$.)

In $^4{\rm He}^+$ measurements of sufficient precision for at least three
transitions seem possible in principle as per above: 1 Hz accuracy has
already been suggested for $1S-2S$ \cite{Herrmann2009}, and we assume the
missing 3 orders of magnitude improvement in $2^3S-2^3P$ to be possible.
These two transitions have nonvanishing Higgs terms which differ in
$\Delta_i$ (Eq. (\ref{eqRx})). Theoretical $^4{\rm He}$ - $^3{\rm He}$
isotope shifts (with the precision required for extracting the charge radii)
for both the $2^3S-2^3P$ and for the $2^1S-2^3S$ transition at $0.8$ eV have
been calculated \cite{PachuckiHeIS2015}. The Higgs contribution vanishes in
the latter but it provides the third equation (\ref{eqdei}), making it
possible to eliminate the nuclear structure terms. It has been measured to
$8 \times 10^{-12}$ or $1.8$ kHz \cite{Rooij2011}. Like in muonic
experiments, current precision is geared to the extraction of charge radii
from isotope shifts.

The denominator of Eq. (\ref{eqdei}) of course vanishes in the leading order
in $(\alpha Z)$. The $A$ dependence of ${\cal B}_i$ starts at order $(\alpha
Z)^6$.

\section{Conclusion}

Current upper bounds on both the muon-nucleon and electron-nucleon Higgs
coupling constrain the possibility of Higgs term extraction from the Lamb
shift to muonic helium and heavier systems. The $Z^4$ scaling of the nuclear
structure corrections, existing experimental work on light muonic systems,
as well as more difficult control of the number of ejected electrons during
the muon cascade in heavier systems, all favor lighter systems, making the
muonic and electronic helium ion the prferred system. Direct extraction of
the Higgs term is not viable because of too large nuclear structure terms,
which exhibit uncertainties too large by $3-4$ orders of magnitude. Due to
their dependence on either the effective nuclear potentials or on assumed
charge distributions, this uncertainty cannot be reduced. Instead, (i)
availability of several transitions is required to eliminate these terms,
and (ii) uncertainty has to be reduced a few orders of magnitude below the
current which is geared towards extracting the charge radii. In muonic
helium the Lamb shift experimental precision should be at least $0.1$ ppm,
possibly requiring the evaluation of small effects \cite{Amaro2015}, but
there is no other suitable, precisely measurable transition. The normal
(electronic) helium ion is more promising, offering the lowest-lying
transition measurable to 1 Hz accuracy and several states currently measured
to kHz precision. We give required elevated precision (above that of the
Rydberg constant) which may allow elimination of nuclear structure terms.
The resolution of the charge radius puzzle which is the current focus of the
light muonic and electronic ion physics does not require this level of
precision.

\section*{Acknowledgments}

I thank S. Fajfer for suggesting to look for atomic Higgs effects in
precisely solvable systems, J. F. Kamenik for discussions and Nir Barnea for
comments on the state of the art {\sl ab-initio} nuclear polarization and
charge radius calculations.


\begin{thebibliography}{99}



\bibitem{P1} C. Delaunay, R. Ozeri, G. Perez and Y. Soreq,
arXiv:1601.05087v1 [hep-ph] (2016).

\bibitem{P1a} C. Delaunay and Y. Soreq,
arXiv:1602.04838v1 [hep-ph] (2016).

\bibitem{P2} C. Frugiuele, E. Fuchs, G. Perez and M. Schlaffer,
arXiv:1602.04822v1 [hep-ph] (2016).



\bibitem{K1} W.H. King,
J. Opt. Soc. Am. {\bf 53}, 638 (1963).

\bibitem{KMmuHe} R. Krivec and V.B. Mandelzweig,
Phys. Rev. A {\bf 56}, 3614 (1997), Phys. Rev. A {\bf 57}, 4976 (1998).


\bibitem{B1} E. Borie and G.A. Rinker,
Rev. Mod. Phys. {\bf 54}, 67 (1982).

\bibitem{CREMA} A. Antognini,
arXiv:1512.01765v2 [physics.atom-ph] (2016).

\bibitem{Lamb2012} T. Nebel et al.,
Hyperfine Int. {\bf 212}, 195 (2012).

\bibitem{CREMA2016} R. Pohl et al.,
arXiv:1609.03440v1 (2016).



\bibitem{FriarFS} J.L. Friar,
Ann. Phys. {\bf 122}, 151 (1979).

\bibitem{B2} E. Borie,
Ann.\ Phys.\ {\bf 327}, 733 (2012); arXiv:1103.1772v7 (2014).

\bibitem{Yerokhin2011} V.A. Yerokhin,
Phy. Rev. A {\bf 83}, 012507 (2011).

\bibitem{Czarnecki2005} A. Czarnecki, U.D. Jentschura and K. Pachucki,
Phys. Rev. Lett. {\bf 95}, 180404 (2005).



\bibitem{Ye1} G. Aad et al. (ATLAS),
Phys. Lett. {\bf 68}, B738 (2014).

\bibitem{Ye2} V. Khachatryan et al. (CMS Collaboration),
Phys. Lett. B {\bf 744}, 184 (2015).

\bibitem{Ye3} W. Altmannshofer, J. Brod and M. Schmaltz,
JHEP {\bf 05}, 125 (2015).

\bibitem{kmu1} {\tt
http://cds.cern.ch/record/2052552/files/ ATLAS-CONF-2015-044.pdf},
last accessed 2016/06/10.



\bibitem{Yq1} M.A. Shifman, A.I. Vainshtein and V.I. Zakharov,
Phys. Lett. B {\bf 78}, 443 (1978).

\bibitem{Yq2} G. Belanger, F. Boudjema, A. Pukhov and A. Semenov,
Comput. Phys. Commun. {\bf 180}, 747 (2009).

\bibitem{Yq3} P. Junnarkar and A. Walker-Loud,
Phys. Rev. D {\bf 87}, 114510 (2013).

\bibitem{Yq4} G. Belanger, F. Boudjema, A. Pukhov and A. Semenov,
Comput. Phys. Commun. {\bf 185}, 960 (2014).

\bibitem{Yq5} ATLAS and CMS (2015),
ATLAS-CONF-2015-044.



\bibitem{Yudsc1} V. Khachatryan et al. (CMS Collaboration),
Eur. Phys. C {\bf 75}, 212 (2015).

\bibitem{Yudsc2} G. Perez, Y. Soreq, E. Stamou and K. Tobioka,
Phys. Rev. D {\bf 92}, 033016 (2015).

\bibitem{Yudsc3} Y. Zhou,
arXiv:1505.06369 (2015).

\bibitem{Yudsc4} C. Delaunay, T. Golling, G. Perez and Y. Soreq,
Phys. Rev. D {\bf 89}, 033014 (2014).

\bibitem{Yudsc5} A.L. Kagan, G. Perez, F. Petriello, Y. Soreq, S.
Stoynev and J. Zupan,
Phys. Rev. Lett. {\bf 114}, 101802 (2015).

\bibitem{Yudsc6} G. Perez, Y. Soreq, E. Stamou and K. Tobioka,
Phys. Rev. D {\bf 93}, 013001 (2016).



\bibitem{Martynenko07} A.P. Martynenko,
Phys. Rev. A {\bf 76}, 012505 (2007).




%



\bibitem{Herrmann2009} M. Herrmann et al.,
Phys. Rev. A {\bf 79}, 052505 (2009).

\bibitem{PastorHeIS2012} P. Cancio Pastor et al.,
Phys. Rev. Lett. {\bf 108}, 143001 (2012).



\bibitem{Carlson2011} C.E. Carlson, V. Nazaryan and K. Griffioen,
Phys. Rev. A {\bf 83}, 042509 (2011).

\bibitem{Carlson2015} C.E. Carlson,
Progress in Particle and Nuclear Physics {\bf 82}, 59 (2015).

\bibitem{Carboni77} G. Carboni et al.,
Nucl. Phys. A {\bf 278}, 38 (1977).

\bibitem{Carboni78} G. Carboni et al.,
Phys. Lett. B {\bf 73}, 229 (1978).

\bibitem{Sick76} I. Sick, J.S. McCarthy and R.R. Whitney,
Phys. Lett. B {\bf 64}, 33 (1976).



\bibitem{Friar03} J.L. Friar,
Lect. Notes Phys. {\bf 627}, 59 (2003).

\bibitem{Sick82} I. Sick,
Phys. Lett. B {\bf 116}, 212 1982.

\bibitem{PRP1} R. Pohl, R. Gilman, G.A. Miller and K. Pachucki,
Ann. Rev. Nucl. Part. Sci. {\bf 63}, 175 (2013); arXiv:1301.0905.

\bibitem{PMP1} T.P. Gorringe and D.W. Hertzog,
Progress in Particle and Nuclear Physics {\bf 84}, 73 (2015).

\bibitem{Bernauer1} J.C. Bernauer et al.,
Phys. Rev. Letters {\bf 105}, 242001 (2010).



\bibitem{DRP1} R. Pohl et al. (CREMA collaboration),
Science {\bf 353}, 669 (2016).

\bibitem{AntogniniPRP1} A. Antognini et al.,
Can. J. Phys. {\bf 89}, 47 (2011).

\bibitem{SickRe4he} I. Sick,
Phys. Rev. C {\bf 77}, 041302 (2008).

\bibitem{Pohl2016a} R. Pohl for the CREMA collaboration,
Journal of the Physical Society of Japan {\bf 85}, 091003 (2016).

\bibitem{Antognini15} A. Antognini et al., arXiv:1509.03235v2 (2015); Proc.
21st International Conference on Few-Body Problems in Physics, Chicago, USA,
May 18-22, 2015, C. Elster, D.R. Phillips and C.D. Roberts (eds.).



\bibitem{Gardner82} C.J. Gardner et al.,
Phys. Rev. Lett. {\bf 48}, 1168 (1982).



\bibitem{Barnea13a} C. Ji, N. Nevo Dinur, S. Bacca and N. Barnea,
Phys. Rev. Letters {\bf 111}, 143402 (2013).

\bibitem{FriarZem} J.L. Friar,
arXiv:1306.3269, Phys. Rev. C {\bf 88}, 034003 (2013).

\bibitem{PachuckiZem} K. Pachucki,
Phys. Rev. Lett. {\bf 106}, 193007 (2011).



\bibitem{PohlSOP} R. Pohl, A. Antognini, F. Nez, F.D. Amaro, F. Biraben et al.,
Nature (and Supplementary Material) {\bf 466}, 213 (2010).

\bibitem{Karshenboim2010} S.G. Karshenboim, V.G. Ivanov, E.Yu. Korzinin and
V. A. Shelyuto,
Phys. Rev. A {\bf 81}, 060501 (2010).

\bibitem{Karshenboim2012} S.G. Karshenboim, V.G. Ivanov and E.Yu. Korzinin,
Phys. Rev. A {\bf 85}, 032509 (2012).

\bibitem{Jentschura2011b} U.D. Jentschura,
Phys. Rev. A {\bf 84}, 012505 (2011).

\bibitem{Elekina2011} E.N. Elekina, A.A. Krutov and A.P. Martynenko,
Phys. Part. Nucl. {\bf 8}, 331 (2011).

\bibitem{Krutov2015} A.A. Krutov et al.,
JETP Lett. {\bf 120}, 73 (2015).



\bibitem{DiepoldPhD} M. Diepold,
PhD Dissertation,
Ludwig-Maximilians-Universit\"{a}t M\"{u}nchen, Munich 2015.



\bibitem{Barnea15b} Chen Ji, O.J. Hernandez, N. Nevo Dinur, S. Bacca and N. Barnea,
arXiv:1509.01430v1 (2016).

\bibitem{Barnea16b} O.J. Hernandez, N. Nevo Dinur, Chen Ji, S. Bacca and N. Barnea,
arXiv:1604.06496v1 (2016).



\bibitem{muHNonPert2013} P. Indelicato,
Phys. Rev. A {\bf 87}, 022501 (2013).

\bibitem{muHNonPert2011} J.D. Carroll, A.W. Thomas, J. Rafelski and G.A. Miller,
Phys. Rev. A {\bf 84}, 012506 (2011); arXiv:1104.2971v3.

\bibitem{Beauvoir2000} B. de Beauvoir et al.,
The European Physical J. {\bf D}, 61 (2000).

\bibitem{PachuckiHeIS2015} K. Pachucki and V.A. Yerokhin,
arXiv:1503.07727v2 (2015).

\bibitem{CODATA2012} P.J. Mohr, B.N. Taylor, and D.B. Newell,
Rev. Mod. Phys. {\bf 84}, 1527 (2012).

\bibitem{Yerokhin2006} V.A. Yerokhin, P. Indelicato and V.M. Shabaev,
arXiv:physics/0611265v1 (2006).

\bibitem{Rooij2011} R. van Rooij et al.,
Science {\bf 333}, 196 (2011); arXiv:1105.4974v1.















\bibitem{Amaro2015} P. Amaro et al.,
Phys. Rev. A {\bf 92}, 022514 (2015).


\end{thebibliography}
\end{document}